\documentclass[doublecol]{epl2} 
\usepackage{amssymb}

\title{Experimental study of shear band formation: bifurcation and
  localization} \shorttitle{Experimental study of shear band formation}

\author{T. B. Nguyen \and A. Amon\footnote{axelle.amon@univ-rennes1.fr}}
\shortauthor{T. B. Nguyen \and A. Amon}

\institute{Universit\'e de Rennes 1, Institut de Physique de Rennes
  (UMR UR1-CNRS 6251), B\^{a}t.~11A, Campus de Beaulieu, F-35042
  Rennes, France }


\abstract{We report the experimental observation of the bifurcation at
  the origin of localization of the deformation in a granular material
  submitted to uniaxial compression. We present a quantitative
  characterization of the heterogeneity in the strain field
  repartition allowing to evidence objectively the existence of a
  bifurcation initiating the shear bands formation process. We show
  that this bifurcation is supercritical and has no clear signature on
  the stress-strain curve. At the bifurcation, a symetry breaking
  occurs characterized by the emergence of a well-defined orientation
  corresponding to the Mohr-Coulomb angle. Yet, plasticity is still
  diffuse and the shear band extension is of the order of the sample
  width. While loading proceeds, the shear band narrows until it
  reaches, after the peak of the stress-strain curve, a stationary
  width.}

\begin{document}

\maketitle

\section{Introduction}
Numerous amorphous materials such as foams, colloidal glasses,
granular materials or metallic glasses, tend to present shear banding
when they are sheared, i.e. the concentration of the deformation in a
subpart of the system, so that a cohabitation between an almost solid
part and a fluid-like part is observed~\cite{schall2010}.

In the case of granular materials, this localization of the
deformation takes the form of failure planes of a few beads diameter
thickness~\cite{davis,vardoulakis,desrues2004}. Because this
localization of the deformation has obvious connection with soils
stability and fault formation, numerous works have been devoted to its
prediction. In the context of elastoplasticity, localization is
described as a bifurcation
phenomenon~\cite{rudnicki1975,bardet1990,vardoulakis} which
corresponds to the emergence of solutions presenting discontinuities
in the strain rate field. The condition yields solely a direction,
without any prediction concerning the number of bands, their thickness
or their position. Experimentally, setup allowing a full-field
observation of the strain repartition during the loading are
scarce~\cite{desrues2004,rechenmacher2006,hall2010a} and there has
been a long debate if shear bands emerge before, at or after the peak
of the loading curve. Recent progresses in the detection
methods~\cite{ando2013,erpelding2013} have allowed to observe
intermittent inhomogeneities in the strain field occuring before
failure~\cite{lebouil2014a,lebouil2014b,desrues2015}. Those results
have renewed the interest for this field of research with a particular
interest for the understanding the microscale processes at the origin
of the shear bands formation~\cite{tordesillas2015}. All those studies
show that the localization process initiates well before the peak of
the loading curve and cannot be identified has the sudden propagation
of a failure plane from a defect. Surprisingly, a clear experimental
signature of the bifurcation coining the initiation of the
localization process is still missing.

As granular materials are amorphous materials, an approach to
understand the physics underlying the shear localization process is to
tackle the problem from the soft glassy materials point of
view. Indeed, recent progresses in the understanding of the elementary
mechanisms at the origin of their plastic response have been
made~\cite{barrat2011} and the similarities in their behaviors allow
to hope for a universal description of the plastic flow of amorphous
systems. Failure and shear bands formation in this framework are
supposed to originate from the elastic long-range coupling between the
elementary plastic events which initiates avalanches of correlated
rearrangements. Corresponding to this picture, recent numerical and
theoretical works show that the failure of amorphous materials could
be seen as a critical
phenomenon~\cite{gimbert2013,lin2015}. The question of the
  nature of this transition and of the universality class to which it
  belongs is one of the key question of the field. Several recent
  works, both experimental~\cite{Dijksman2011,Chikkadi2014} and
  theoretical~\cite{Jaiswal2016} point towards a discontinuous phase
  transition, i.e. a first-order transition. Practically, depending
on the material and on the loading conditions, the precise form and
nature of the shear bands differ from one system to
another~\cite{schall2010}. It is thus not clear if the universality of
the mechanism of plasticity at the elementary scale hold when the
plasticity self-organizes at a larger scale into shear bands.

Here we present an experimental study of the whole process of failure
in a granular sample submitted to a biaxial test, from the initial
homogeneous deformation to final permanent shear bands. Using image
analysis tools, we quantify the anisotropy in the spatial distribution
of the strain and exhibit objective quantities to characterize shear
bands formation. We demonstrate experimentally the existence of a
bifurcation during the loading process corresponding to a symetry
breaking and to the birth of a definite direction in the strain
field. We show that this bifurcation is supercritical: no
discontinuity occurs in the observed field during the loading and at
the bifurcation, the strain field is still diffuse in the material. As
the loading proceed, the thickness of the shear band decreases until
it reaches a stationary value.

In a first part, we describe the experimental setup. In a second part
we present the image analysis tools which allow us to give an
objective measurement of the spatial repartion of the deformation in
the sample and we apply it to the analysis of our experiments. In the
last part, we discuss our results in the framework of different
theories.

\section{Experimental setup}
\begin{figure}[htbp]
\centerline{\includegraphics[width=\linewidth]{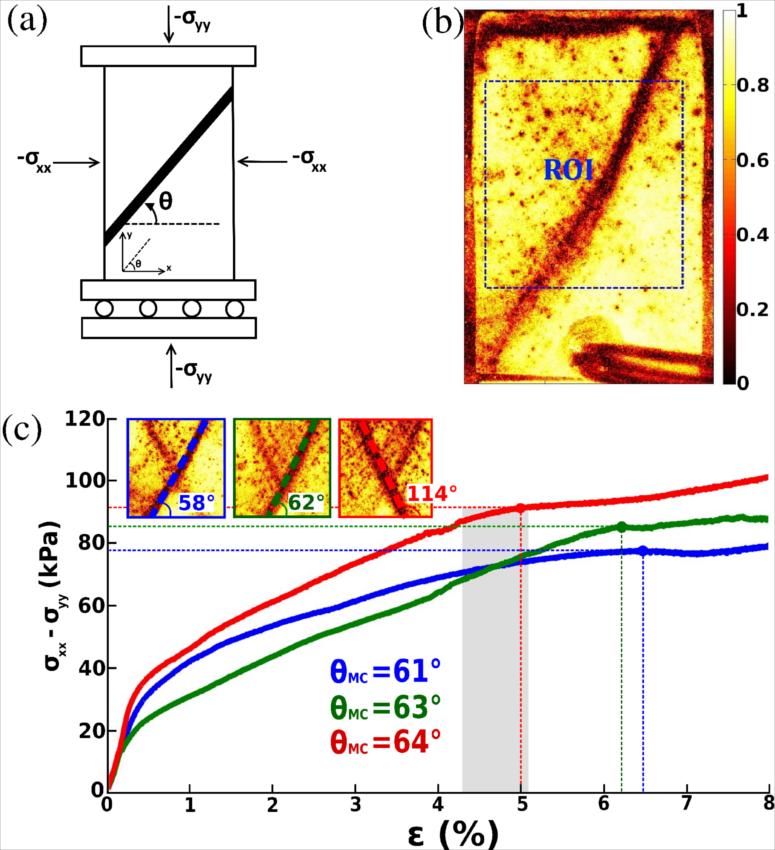}}
\caption{(a) Schematic of the biaxial test. (b) Example of a
  correlation map with the color code scale. The square region of
  interest (ROI) used for image analysis is shown in blue. (c) Loading
  curves for three different runs. The gray zone is the same as in
  Fig~\ref{fig:anisotropy} and indicates roughly the position of the
  bifurcation for all three experiments. The values of the
  Mohr-Coulomb angles deduced from the loading curves are given for
  each run ($\theta_{MC}$). Upper insets: correlation maps at the
  plateaus with direct measurement of the shear bands inclination.}
\label{fig:setup}
\end{figure}

The setup is a biaxial compressive test extensively described in
another publication~\cite{lebouil2014a} and schematized in
Figure~\ref{fig:setup}. The material consists in dry glass beads of
diameter $d = 70-110~\mu$m. It is placed between a preformed latex
membrane (85 $\times$ 55 $\times$ 25 mm$^3$) and a glass plate. A pump
produces a partial vacuum inside the membrane, creating a confining
pressure $-\sigma_{xx}$. The preparation of the sample ensures
reproducible experiments at a volume fraction of $\approx 0.60$. The
prismatic sample thus obtained is placed in a testing machine which
enforces the following conditions: (i) the back plate and the front
glass plate forbid displacements normal to their plane and thus ensure
plane strain conditions; (ii) a roller bearing at the bottom allows
for a translational degree of freedom. This feature is a modification
of the setup of ref.~\cite{lebouil2014a} which allows a breaking of
symmetry when failure occurs as we will see in the following; (iii)
the upper plate is displaced vertically by a stepper motor with a
velocity of $1~\mu$m/s leading to a deformation rate of $1.1 \times
10^{-5}$ s$^{-1}$. A sensor fixed on the top plate measures the force
exerted in the $y$ direction from which the stress on the plate is
deduced. The value of the confining stress for all the experiments
presented here is 30~kPa so that crushing of particles is not
expected. The loading curves of three different experiments prepared
in the same conditions are shown in Figure~\ref{fig:setup}(c). Such
curves are very similar to loading curves reported in the literature
for not too dense samples. A plateau with a local maximum can be
identified for each of those curves indicating a modification in the
response of the material. Those changes in response are usually
identified as indicator of the failure of the material: after this
point the sample is supposed to be best represented as separated in
blocks in relative solid translation. In our experiments those local
maxima occur at a value of the deviatoric stress in the range 80 to
90~kPa, corresponding to an axial deformation of 5 to 6.5~\%. From the
value of stress at those points, the Mohr-Coulomb angle,
$\theta_{MC}$, can be deduced for each experiment~\cite{nedderman}:
$\theta_{MC} = 45^\circ + \frac{1}{2} \sin^{-1} \left[
    \frac{\sigma_{yy} - \sigma_{xx}}{\sigma_{yy} +
      \sigma_{xx}}\right]$. Our measurements give an average of
$62.5^\circ \pm 1.5^\circ$.

\begin{figure*}[htbp]
\includegraphics[width=\linewidth]{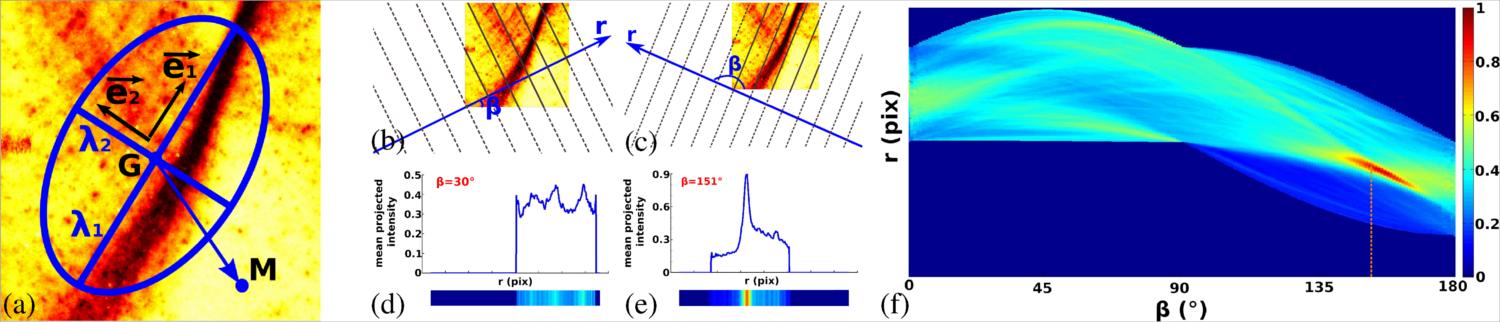}
\caption{(a) Graphical representation of the inertia tensor
  $\overline{\overline{P}}$ computed from the pixels of the image
  using eq.~\ref{eq:tensor}. (b) to (e): Principle of the projection
  method. (b) and (c) The intensities of the pixels are averaged along
  the direction orthogonal to a line given by its slope $\beta$ ((b)
  $\beta = 30^{\circ}$, (c) $\beta = 151^{\circ}$). (d) and (e)
  Unidirectional curve of averaged intensity obtained for each value
  of $\beta$. Those curves can also be represented using a colorscale
  as shown underneath each curve. (f) Juxtaposition of the projected
  profiles obtained for all values of $\beta$. From the value of the
  maximum of the map, the principal orientation in the initial image
  can be deduced.}
\label{fig:method}
\end{figure*}
We use a full-field method based on dynamic light scattering for the
quantification of the spatial repartition of the deformation. This
method has been described in details in previous
publications~\cite{erpelding2008,erpelding2013}. A laser beam (532 nm)
is expanded to illuminate the material. The light is multiply
scattered inside the sample and we collect the backscattered
light. The multiple light rays interfer to form a speckle pattern. The
image of the front side of the sample is recorded by a 7360 $\times$
4912 pixels camera. Images are subdivided in square zones of size 16
$\times$ 16 pixels. For each zone the correlation between two
successive speckle patterns $1$ and $2$ is computed as:
\begin{equation}
g_I^{(1,2)} = \frac{\langle I_1 I_2 \rangle - \langle I_1 \rangle
  \langle I_2 \rangle}{\sqrt{\langle I_1^2 \rangle - \langle I_1
    \rangle^2} \sqrt{\langle I_2^2 \rangle - \langle I_2
    \rangle^2}}
\end{equation}
where $I_1$ and $I_2$ are intensities of the pixels of two successive
images and the averages $\langle \cdot \rangle$ are done over the 16
$\times$ 16 pixels of a zone. Each computed value gives a pixel in the
final correlation map and corresponds to a volume of area in the front
plane $2 d \times 2 d$ and depth of a few $d$. An example of a map of
correlation is shown in Fig.~\ref{fig:setup}(b). The normalisation of
the correlation function ensures that the correlation values are in
the interval $[0,1]$ and the colorscale used in all the following is
shown in Figure~\ref{fig:setup}(b). The decorrelation of the scattered
light comes from relative beads motions, i.e. combination of affine
and nonaffine bead displacements and rotation of nonspherical
beads. In the following, the axial deformation increment between the
two images used to obtain a map is 3.2 $\times 10^{-5}$.

\section{Spatial repartition of the deformation}
A movie of the successive correlation maps during the loading is in
supplemental material. The general phenomelogy observed during the
loading has been already described
elsewhere~\cite{lebouil2014a,lebouil2014b}. Strain repartition in the
sample is inhomogeneous. Intermittent micro-bands are observed from
the beginning of the loading until the final shear bands are
established. This microstructure has been studied before and the
orientation of the micro-bands differ from the orientation of the
final shear bands\cite{lebouil2014b}. The former corresponds to the
orientation predicted by elasticity for stress released by local
plastic rearrangements (Eshelby's quadrupolar
redistribution~\cite{eshelby1957}) and does not depend of the
frictional properties of the material. On the other hand distinct
final shear bands are observed in all our experiments with an
orientation in agreement with the Mohr-Coulomb angle deduce from the
loading curves~\cite{nedderman} as is shown in insets of
Fig.~\ref{fig:setup}(c).

The goal of the present letter is to study the process of formation of
the final shear bands. For our purpose, we need to exhibit objective
quantities from our deformation maps. To remove intermittent
fluctuations, we average stacks of 50 consecutive maps to obtain a
smooth strain field. We underline that no further image treatment is
done on the images. In the following, we first present the image
analysis tools we use to characterize the degree of localisation in
the sample and the orientation of the observed large-scale
structure. In a second part, we use those tools to study shear band
formation.

\subsection{Image analysis}
To quantify the emergence of an oriented structure at a large scale
and its degree of localization, we use two different tools that we
call in the following anisotropy measurement for the first one and
projection analysis for the second one. For those two methods we need
that information encoded in the image (i.e. deformation) have the
largest weight when the correlation is low. Consequently we define for
each correlation map its invert image which we call activity
image. The values of the pixels of those images are given by
$I_A(\vec{r}) = 1 - g_I(\vec{r})$ and all the subsequent analyses are
done on the activity images.

To characterize the degree of anisotropy in an image we use a method
extensively described elsewhere~\cite{lehoucq2015}. Considering the
value of each pixel of an activity image as a weight, we compute the
center of mass of the image $G$ and the inertia tensor
$\overline{\overline{P}}$ which quantify the spatial repartition of
the weights around $G$ (see Fig.~\ref{fig:method}(a)):
\begin{equation}
\overline{\overline{P}} = \sum_{\vec{M}} I_A(\vec{M}) \overrightarrow{GM} \otimes
\overrightarrow{GM}
\label{eq:tensor}
\end{equation}
This tensor has two positive eigenvalues noted $\lambda_1 > \lambda_2
> 0$. The eigenvector associated with the largest eigenvalue gives the
principal direction of the anisotropy in the image and the anisotropy
index $a=1 - \frac{\lambda_2}{\lambda_1}$ gives a measurement of the
degree of anisotropy between 0 (isotropic) and 1 (maximal
anisotropy). A visualization of the tensor can be given by an ellipse
of axis determined by the eigenvalues and eigenvectors of
$\overline{\overline{P}}$ (see Fig.~\ref{fig:method}(a)).

The second method consists in projecting the intensity of the activity
image on a line going through the origin and with an angle $\beta$
with the $x-$axis~\cite{welker} (see Fig.~\ref{fig:method}(b) and
(c)). Different values of $\beta$ provide different projection
profiles of the same image as shown in Fig.~\ref{fig:method}(d) and
(e) which are the profiles obtained for respectively
Figure~\ref{fig:method}(b) and Figure~\ref{fig:method}(c). The
aggregation of all the profiles obtained for all possible values of
$\beta$ are shown in Figure~\ref{fig:method}(f). The line for which
the profile present a maximum provides an orientation of the structure
(orthogonal to the line) and the full width at half maximum (FWHM) of
the profile provides a width of the structure. Note that this method
is very close in spirit to the Hough transform~\cite{duda1972}.

\subsection{Results}
\begin{figure}[ht]
\includegraphics[width=\linewidth]{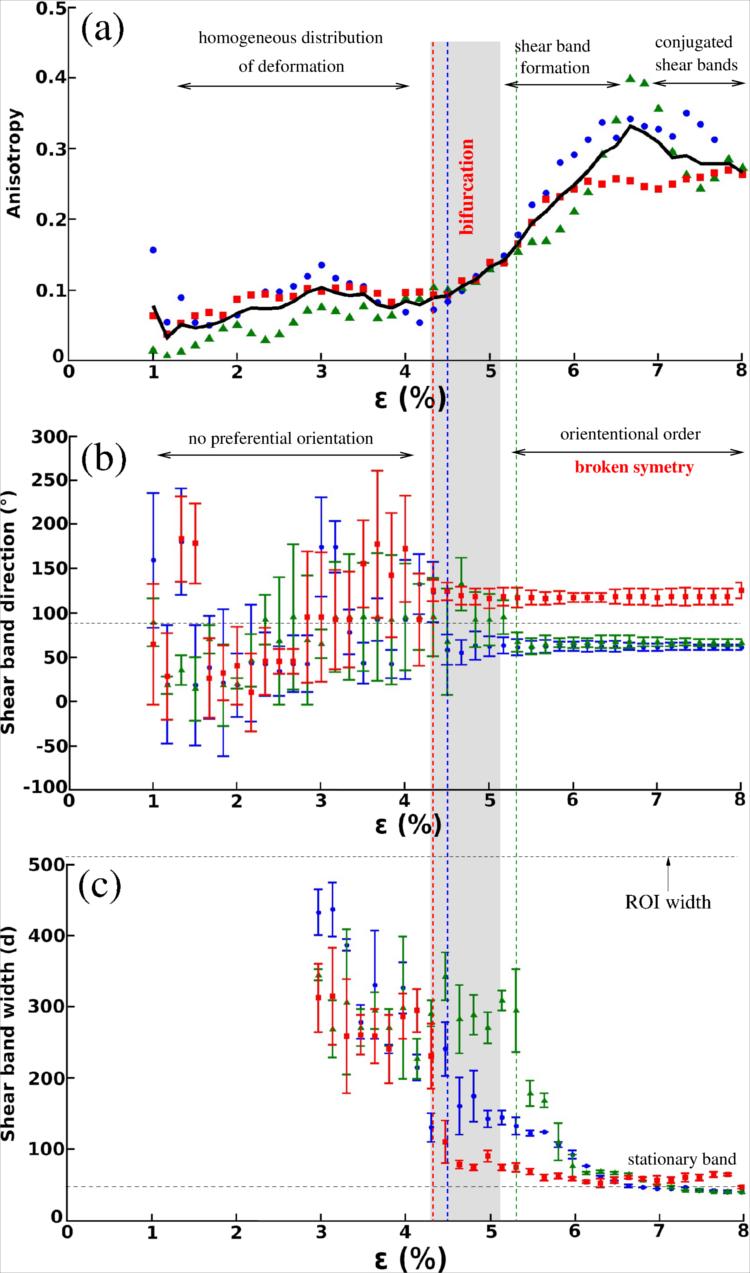}
\caption{(a) Anisotropy of the spatial distribution of the deformation
  as a function of the axial deformation. Data points correspond to
  the three experiments of Fig.~\ref{fig:setup}(c). The black solid
  line is the average of those measurements. (b) Principal orientation of
  the deformation as a function of the axial deformation for the three
  same experiments. Horizontal dashed line: $90^\circ$. Vertical
  dashed lines indicate the axial strain from which the orientation
  has been unambiguously chosen (drastic decrease of the uncertainty)
  for each experiment. The shadow box has been traced from the mean
  and standard deviation of those values of the axial strain ($4.7 \pm
  .4$~\%). (c) Thickness of the shear band rescaled by the mean
  diameter of the beads as a function of the axial deformation.}
\label{fig:anisotropy}
\end{figure}

Figure~\ref{fig:anisotropy}(a) shows the anisotropy of the spatial
distribution of the deformation as a function of the axial deformation
$\epsilon$ for three different experiments (same color code as
Fig.~\ref{fig:setup}(c)). We observe a good reproducibility from one
experiment to the other. The black curve is the mean anisotropy
obtained by an average over the different experiments. At the very
beginning of the loading ($\epsilon \lesssim 1$\%), the settlement of
the sample leads to spurious heterogeneous effects that leads to
unrelevant large anisotropy values (see movie in supplemental
material). For axial deformations between 1\% and about 4.5\% the
anisotropy is low (about 0.1) and the spatial repartition of the
strain can be considered as isotropical on average. From a value of
the axial deformation of about 4.5\%, the anisotropy increases
steadily until it reaches a maximal value at about 6.8\%. Beyond this
value, depending on the experiment, the anisotropy stays for further
loading roughly constant or decreases. Decreases are due to the
emergence of a secondary shear band, conjugated to the initial one
(see movie in supplemental material).

Figure~\ref{fig:anisotropy}(b) shows the principal orientation of the
strain distribution (projection analysis). For an axial deformation
smaller than $\sim$4.5\%, the direction obtained are randomly
distributed and the errors on the determination of the angles are
large, indicating that no clear orientation can be defined in the
images. Beyond $\epsilon \simeq 4.5$~\%, each experiment displays a
well-defined, constant, orientation. Depending on the experiment this
direction is either about $60^\circ$ or $120^\circ$, but the ensemble
of orientations collected on all the experiments are symetrical
compared to $90^\circ$. There has been thus a symetry breaking between
two possible directions. For one of the experiments (green triangles),
the uncertainty on the angle stays large until $\epsilon \simeq
5.3$~\%. Still, its anisotropy is increasing steadily from $\epsilon
\simeq 4.5$~\% showing that an orientation has emerged even if the
band is less clearly defined compared to the other two experiments.

Figure~\ref{fig:anisotropy}(c) displays the evolution of the FWHM of
the profile obtained along the principal orientation of the spatial
distribution of the strain using the projection method. For values of
the axial strain smaller than 4~\%, the projected profiles are mainly
flat with noise so that thickness measurements lead to noisy values of
the order of size of the ROI. Thus, FWHM are clearly defined only when
the orientation have been determined with small uncertainty. We have
added an indication of the axial deformation for which the orientation
is well-defined for each run (vertical dashed lines). We observe that
for all the experiments, the width of the band is still very large (of
the order of half the size of the ROI) even though the orientation is
unambiguously established and that we are sure that the bifurcation
has been crossed. For further loading, a decrease of the width is
observed until a stationary value is reached. This stationary value is
about 50$d$ and stays roughly constant for further loading (i.e. above
8\%).


\section{Discussion and conclusion}
The first result that emerges from Fig.~\ref{fig:anisotropy} is that
localization of deformation is indeed initiated by a bifurcation that
occurs at an axial strain of about $\epsilon \simeq 4.5$~\%. The
observed bifurcation coins a breaking of symetry in the distribution
of the strain (see Fig.~\ref{fig:anisotropy}(b)) simultaneous to an
increase of the anisotropy of this strain distribution (see
Fig.~\ref{fig:anisotropy}(a)). This bifurcation does not correspond to
a sudden localization of the strain distribution in the form of a
failure plane. On the contrary, the deformation is still largely
diffuse in the sample at the bifurcation (see
Fig.~\ref{fig:anisotropy}(c)). This bifurcation is thus
\emph{supercritical}: neither the anisotropy
(Fig.~\ref{fig:anisotropy}(a)), nor the width of the band
(Fig.~\ref{fig:anisotropy}(c)) present a discontinuity. This
experimental observation confirms the prediction of several theories
that the bifurcation corresponds to the emergence of a direction
without any definition of a particular plane or a finite thickness. It
is interesting to note that this supercritical feature is closely
linked to the progressive organization of plasticity in the material
and hence to the ductile nature of the failure.

The location of the bifurcation is reported in Fig.~\ref{fig:setup}(c)
as a shaded region. It can be seen that this bifurcation occurs well
before the plateaus of the loading curves except for one of the
experiment. It has to be noted that this particular experiment (red
squares) is the one that localizes the first in
Fig.~\ref{fig:anisotropy}(b) so that the bifurcation also happened
before the maximum in this experiment. After the bifurcation, the
finite direction that has emerged stays constant for further loading
as can be seen in Fig.~\ref{fig:anisotropy}(b). Its value corresponds
to the Mohr-Coulomb angle which can be deduced from the plateaus of
each stress-strain curves. This result is surprising: the Mohr-Coulomb
angle emerges in the material \emph{before} the yield stress is
reached and the internal friction manifests itself while the system
cannot be represented as two blocks sliding one against the other as a
rough picture of the Mohr-Coulomb model would
suggest~\cite{nedderman}.

After the bifurcation, the band is progressively forming as can be
characterized by the decrease of its thickness (see
Fig.~\ref{fig:anisotropy}(c)). It reaches a stationary value at an
axial strain of about $\epsilon \gtrsim 6.5$~\%. The shear band
thickness we measure ($\sim 50 d$) is significantly larger that the
ones reported in the litterature~\cite{vardoulakis,rechenmacher2006}
by more direct measurements which are generally about $\sim 10-20
d$. This is due to the high sensitivity in strain of our measurement
method. Indeed, it is now well-established that the transition between
the liquid-like and solid-like regions of a granular material are not
well-defined~\cite{komatsu2001} and that creeping flow with an
exponential decay of the velocity can be detected even very far from a
shear band (see e.g.~\cite{crassous2008} and references therein). In
fact, recent experiments show that in presence of a shear band the
whole material is flowing and no solid part can be properly
defined~\cite{nichol2010,reddy2011}. Because our method is able to
detect deformation of order $10^{-5}$~\cite{lebouil2014a}, the shear
bands we observe in our correlation maps are larger than the ones
obtained by more direct measurement.

Bifurcation theories that describe failure in soil
mechanics~\cite{rudnicki1975,vardoulakis} rely on the assumption of
the existence of discontinuous solutions of the strain rate field. As
experimentally such discontinuities are not observed, those models are
not fully satisfying to describe the behavior of the system. The need
to describe the strain field during localization as a smooth
inhomogeneous field calls for new models.

Nonlocal rheological models inspired by theoretical works on soft
glassy materials have been proposed to describe the rheology of
granular materials~\cite{kamrin2012,bouzid2015}. Those models belong
to continuum mechanics and introduce a new variable to describe the
local state of the material, the \emph{fluidity}. Spatial
heterogeneities in the flow emerge because this fluidity obey to a
partial differential equation (PDE) coupled to the constitutive
law. This PDE describes the spatial dynamics of the fluidity by a
diffusive term and introduces a lengthscale, the ``cooperative
length'' which roughly represents the extension of the perturbation
caused by a flowing zone (see reference~\cite{bouzid2015} for a
discussion of the differences between the models). Such models are
able to describe heterogeneous dense flows in numerous configurations
as well as slow flows far from the shear
bands~\cite{henann2014}. Still, they have been developed to describe
steady flows, in particular they don't consider any volume change, and
it is unclear if they could describe the process of failure, which is,
by nature, a transient. It could seem unlikely that a partial
differential equation including a diffusive term for the ability to
flow (Laplacian of the fluidity) could describe the spontaneous
narrowing of the band after the bifurcation. In fact, dispersive
effects can be cancelled by non-linear effects, leading to
soliton-like solution as has been shown in a very recent theoretical
work~\cite{benzi2016}. Shear bands could then be seen as localized
solution in the dynamical systems meaning of the term. Nevertheless, a
clear physical picture of the competing effects at play in the process
is still missing, which points out the necessity of a clear
understanding of what is hidding behind the fluidity state variable.

To conclude, we have shown experimentally that the failure of a
granular material in uniaxial compression is linked to a supercritical
bifurcation. This bifurcation coined the emergence of a definite
orientation in the diffuse plasticity field. The orientation observed
corresponds to the Mohr-Coulomb angle deduced from the plateau of the
loading curve but its emergence preceeds this plateau. A strong
narrowing of the band follows as the loading proceed until a
stationary size is reached. In the present work, the fluctuations of
the strain field have been removed by the averaging process that
allows image analysis. Those fluctuations in the vicinity of the
bifurcation are awaited to play a major
role~\cite{gimbert2013,lin2015} and are probably linked to the
micro-bands observed at the earlier stages of the
loading~\cite{lebouil2014b}. The study of those fluctuations is the
subject of a future work.

\acknowledgments
It is a pleasure to thank J. Crassous, S. McNamara and J. Weiss for
fruitful discussions.


\begin{thebibliography}{0}
\bibitem{schall2010} Schall, P. and van Hecke, M. {\em
  Annu. Rev. Fluid Mech.} {\bf 42}, 67--88 (2010).

\bibitem{davis} \textit{Plasticity and geomechanics}, Davis, R. O. and
  Selvadurai, A. P. S. (Cambridge University Press, 2002).

\bibitem{vardoulakis} {\em Bifurcation analysis in
  geomechanics}. Vardoulakis, I. and Sulem, J. (Blackie Academic and
  Professional, Glasgow, England, 1995).

\bibitem{desrues2004} Desrues, J. and Viggiani, G. {\em
  Int. J. Numer. Anal. Methods Geomech.}, {\bf 28}, 279--321 (2004).

\bibitem{rudnicki1975} Rudnicki, J. W. and Rice, J. R. {\em
  J. Mech. Phys. Solids}, {\bf 23}, 371 (1975).

\bibitem{bardet1990} Bardet, J.P. {\em Computers and geotechnics} {\bf
  10}, 163 (1990).
\bibitem{rechenmacher2006} Rechenmacher, A. L. {\em
  J. Mech. Phys. Solids} {\bf 54} 22 (2006).

\bibitem{hall2010a} Hall, S. {\it et al.}  {\em G\'eotechnique} {\bf
  60}, 315 (2010).

\bibitem{ando2013} And\' o, E., Hall, S. A., Viggiani, G. and Desrues,
  J. {\em G\'eotech. Lett.}  {\bf 3}, 142 (2013).

\bibitem{erpelding2013} Erpelding, M. {\it et al.} {\em Strain} {\bf
  49}, 167 (2013).

\bibitem{lebouil2014a} Le Bouil, A. {\it et al.} {\em Gran. Matt.}
  {\bf 16}, 1 (2014).

\bibitem{lebouil2014b} Le Bouil, A., Amon, A., McNamara, S., and
  Crassous, J., {\em Phys. Rev. Lett.} {\bf 112}, 246001 (2014).

\bibitem{desrues2015} Desrues, J. and And\' o, E. {\em C. R. Physique}
  {\bf 16}, 26 (2015).

\bibitem{tordesillas2015} Tordesillas, A. {\it et al.} {\em EPL} {\bf
  110}, 58005 (2015).

\bibitem{barrat2011} Barrat, J.-L. and Lema{\^i}tre, A. in {\em
  Dynamical heterogeneities in glasses, colloids, and granular media}
  (eds Berthier, L. {\it et al.}) Ch. 8, 264--297 (Oxford University
  Press, 2011).

\bibitem{gimbert2013} Gimbert, F., Amitrano, D., and Weiss, J., {\em
  EPL} {\bf 104}, 46001 (2013).

\bibitem{lin2015} Lin, J., Gueudr\'e, T., Rosso, A., and Wyart, M.,
  {\em Phys. Rev. Lett.} {\bf 115}, 168001 (2015).

\bibitem{Dijksman2011} Dijksman, J. A., Wortel, G. H., van Dellen,
  L. T. H., Dauchot, O., and van Hecke, M., {\em Phys. Rev. Lett.}
  {\bf 107}, 108303 (2011).

\bibitem{Chikkadi2014} Chikkadi, V., Miedema, D. M., Dang, M. T.,
  Nienhuis, B., and Schall, P., {\em Phys. Rev. Lett.}
  {\bf 113}, 208301 (2014).

\bibitem{Jaiswal2016} Jaiswal, P. K., Procaccia, I., Rainone, C., and
  Singh, M., {\em Phys. Rev. Lett.} {\bf 116}, 085501 (2016).

\bibitem{nedderman} \textit{Statics and Kinematics of Granular
  Materials}, R. M. Nedderman (Cambridge University Press, 1992).

\bibitem{erpelding2008} Erpelding, M., Amon, A., and Crassous,
  J. {\em Phys. Rev. E} {\bf 78}, 046104 (2008).

\bibitem{eshelby1957} Eshelby, J. D. {\em Proc. R. Soc. A}, {\bf 241},
  376 (1957).

\bibitem{lehoucq2015} Lehoucq, R. {\it et al.} {\em Front. Phys.},
  {\bf 2}, 84 (2015).

\bibitem{welker} Welker, P., PhD Thesis, Universit\"at Stuttgart
  (2011) http://dx.doi.org/10.18419/opus-5019.

\bibitem{duda1972} Duda, R. O., and Hart, P. E., {\em Comm ACM} {\bf
  15}, 11 (1972).

\bibitem{komatsu2001} Komatsu, T. S., Inagaki, S., Nakagawa, N., and
  Nasuno, S., {\em Phys. Rev. Lett.} {\bf 86}, 1757 (2001).

\bibitem{crassous2008} Crassous, J., {\it et al.}, {\em
  J. Stat. Mech. Theory Exp.} {\bf 3} P03009 (2008).

\bibitem{nichol2010} Nichol, K. {\it et al.}, {\em Phys. Rev. Lett.}
  {\bf 104}, 078302 (2010).

\bibitem{reddy2011} Reddy, K. A., Forterre, Y., and Pouliquen, O.,
  {\em Phys. Rev. Lett.} {\bf 106} 108301 (2011).

\bibitem{kamrin2012} Kamrin, K., and Koval, G., \textit{Phys. Rev. Lett.}
  \textbf{108}, 178301 (2012).

\bibitem{bouzid2015} Bouzid, M. {\it et al.}, \textit{Eur. Phys. J. E}
  \textbf{38}, 125 (2015).

\bibitem{henann2014} Henann, D. L., and Kamrin, K.,
  \textit{Phys. Rev. Lett.}  \textbf{113}, 178001 (2014).

\bibitem{benzi2016} Benzi, R. {\it et al.}, \textit{Soft Matter}
  \textbf{12}, 514 (2016).

\end{thebibliography}
\end{document}